\documentclass[twocolumn,amssymb,amsmath,floatfix,superscriptaddress,nofootinbib,aps,prc]{revtex4-2}

\usepackage{epsfig}
\usepackage{graphicx}
\usepackage{float}
\usepackage{hyperref}
\usepackage{amsmath}
\usepackage{amssymb}

\begin{document}
	
\title{ Momentum dependent measures of correlations between mean  transverse momentum and harmonic flow in heavy ion collisions}

\author{Rupam Samanta}
\email{rsamanta@agh.edu.pl}
\affiliation{AGH University of Krakow, Faculty of Physics and
	Applied Computer Science, aleja Mickiewicza 30, 30-059 Cracow, Poland}	

\author{Piotr Bożek}
\email{piotr.bozek@fis.agh.edu.pl}
\affiliation{AGH University of Krakow, Faculty of Physics and
	Applied Computer Science, aleja Mickiewicza 30, 30-059 Cracow, Poland}

\begin{abstract}

  The correlation between the mean transverse momentum and the harmonic flow coefficients is an observable which is of great interest; it is sensitive to shape fluctuations in the initial state of a relativistic nuclear collision. The measurement of that  correlation coefficient in central collisions allows one to infer about the intrinsic deformation of the colliding nuclei. We propose to study the momentum dependent covariance and correlation coefficient between the mean transverse momentum and the harmonic flow in a given transverse momentum bin. Two possible constructions of such observables are provided and predictions are obtained from a viscous hydrodynamic model. We find that such momentum dependent correlation coefficients between the mean transverse momentum and the harmonic flow show a strong and  nontrivial momentum dependence. We also explore the effects of granularity (nucleon width) in the initial state, the nuclear deformation, and  the shear viscosity on this momentum dependent correlation coefficient.  The shape of the momentum dependence of the  correlation coefficient for the triangular flow is found to be sensitive to  the size of small scale fluctuations in the initial state. On the other hand, the shape of the momentum dependence of the  covariance between the mean transverse momentum and the harmonic flow coefficients is found to be sensitive to the value of the shear viscosity and to the granularity of the initial state.
\end{abstract}


\maketitle

\section{Introduction}\label{introduction}

The dynamics of the dense fireball formed in relativistic nuclear collisions can be studied using the collective flow observables extracted from the spectra of emitted particles
\cite{Huovinen:2006jp,Voloshin:2008dg,Hirano:2008hy,Heinz:2009xj,Heinz:2013th}. An important part of such studies is devoted to the understanding of event-by-event fluctuations of the collective flow
\cite{Aguiar:2001ac,Takahashi:2009na,Alver:2010gr,Schenke:2010rr,Schenke:2012wb}. The fluctuations of the  harmonic flow coefficients could be due to fluctuations of the shape of the  initial fireball, as well as due to  dynamical fluctuations in the expansion dynamics. Analogously, fluctuations of the mean transverse momentum of emitted particles can be related to the fluctuations of the size of the fireball \cite{Broniowski:2009fm}. The Pearson correlation coefficient between the mean transverse momentum and the harmonic flow coefficients \cite{Bozek:2016yoj}, $\rho([p_T],v_n^2)$,  has been found to be sensitive to  correlations present in the initial state \cite{Bozek:2020drh,Schenke:2020uqq,Giacalone:2020dln}, and also been used as a tool to infer the deformation parameters of the colliding nuclei \cite{Giacalone:2019pca,Giacalone:2020awm,ATLAS:2021kty,JiaIS2021,Jia:2021qyu,Jia:2021wbq,Bally:2021qys,ALICE:2021gxt,ATLAS:2022dov}.

 Experimental and theoretical analyses  of collective flow observables include also the study of event-by-event fluctuations, e.g., using higher moments or higher cumulants of the measured quantities. Complementary information on the multidimensional probability distribution of the considered set of observables involves the covariances between different observables. For the harmonic flow, the symmetric cumulants can be used as observables measuring cross-correlations between harmonic flow coefficients \cite{Bilandzic:2013kga}. 
The  correlation coefficient, $\rho([p_T],v_n^2)$, measures the correlation between the mean transverse
momentum and the momentum averaged harmonic flow coefficient \cite{Bozek:2016yoj}. The class of observables based on the covariance between harmonic flow observables and the mean transverse momentum can be generalized also to momentum dependent observables. In this paper, we propose to measure the momentum dependent correlation coefficient between the mean transverse momentum in an event and the harmonic flow coefficient in a given transverse momentum bin. 

Besides giving  complementary statistical information of the event-by-event distribution of observables, the study of such momentum dependent correlation coefficients could be potentially helpful in elucidating several interesting issues. There are  several  motivations to engage in  studies of this class of observables. The analysis   of  momentum dependent correlators involving the average transverse momentum and  the  harmonic flow coefficients
\begin{itemize}
\item   could clarify
  the observed dependence of the momentum independent coefficient, $\rho([p_T],v_n^2)$, on the transverse momentum cut \cite{ATLAS:2019pvn},
\item could provide information on  specific modes in the initial state related to the final transverse and harmonic flow \cite{Mazeliauskas:2015efa},
\item could give a measure of the correlation between the transverse momentum and harmonic flow irrespective of the shape of the specific momentum dependence of the harmonic flow,
\item could test a possible dependence of the hadronization mechanism on the transverse expansion, when used for identified particles,
\item could help in identifying correlations between the mean transverse momentum and the harmonic flow from the color glass condensate dynamics \cite{Altinoluk:2020psk},   or
  \item  could be sensitive to  the granularity of the initial state  \cite{Giacalone:2021clp}.
  \end{itemize}

In the next section, we write the possible definitions of the momentum dependent correlation coefficient between the mean transverse momentum and the harmonic flow. We show that the momentum dependent correlation coefficients have a robust, nontrivial   momentum dependence.
In section  \ref{sec:granularity} we show that the momentum dependence of the correlation coefficient is sensitive to the granularity of the initial state of the collision. We consider simplified expressions for the momentum dependent correlation coefficient that might be easier to use in experimental analyses (Sec. \ref{sec:other}). The momentum dependent covariance is discussed as a possible observable in  Sec. \ref{sec:covariance}. The discussion is illustrated by the numerical results obtained from a viscous hydrodynamic model for Pb+Pb collisions at $\sqrt{s_{NN}}=5.02$ TeV and U+U collisions at  $\sqrt{s_{NN}}=193$ GeV.

\section{Momentum dependent correlation coefficient between mean transverse momentum and harmonic flow}
\label{sec:diff}

\begin{figure}
	\vspace{5mm}
	\begin{center}
	  \includegraphics[width=0.4\textwidth]{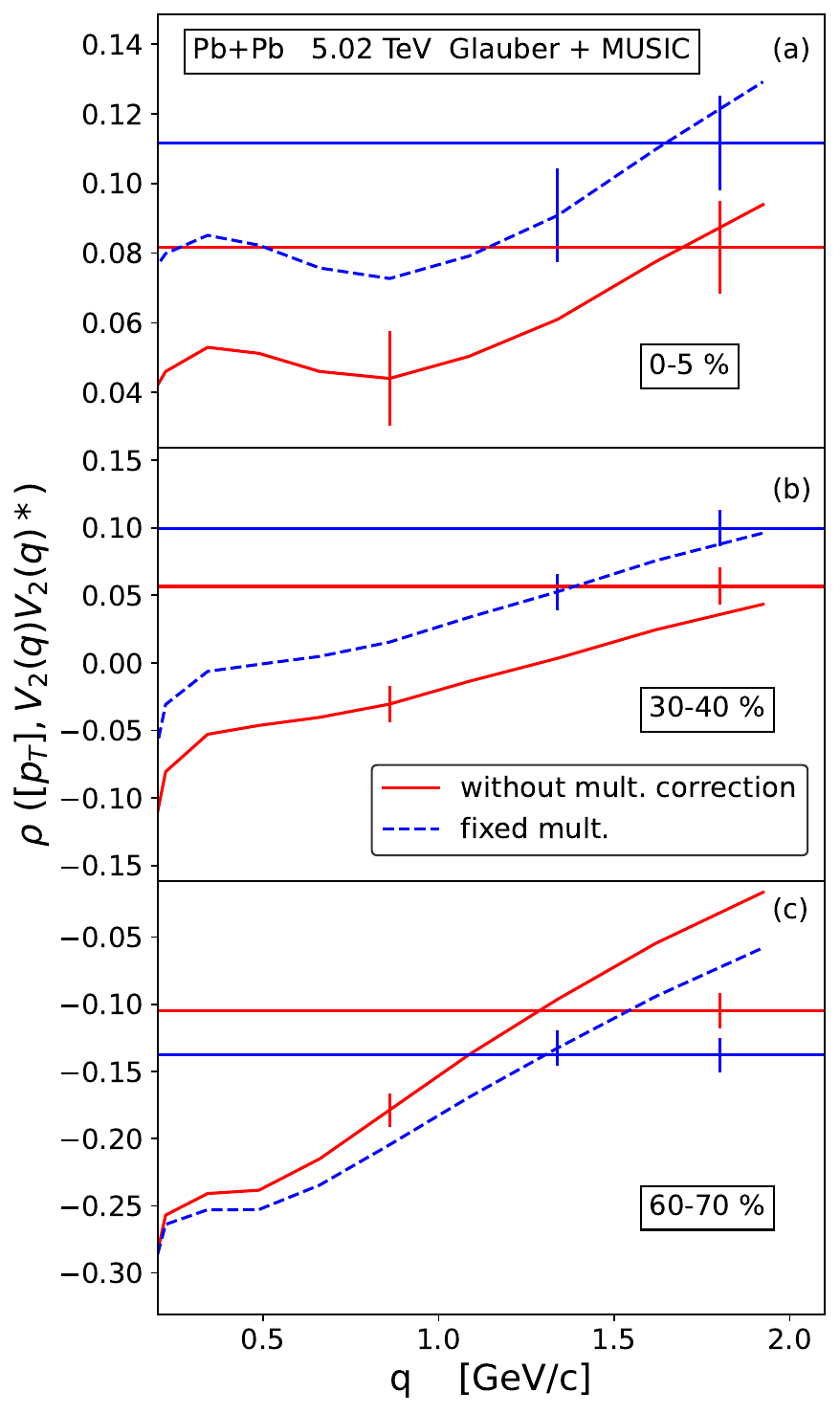} 
	\end{center}
	\caption{The momentum dependent correlation coefficient $\rho([p_T],V_2(q)V_2(q)^\star)$ in Pb+Pb collisions at $\sqrt{s_{NN}}=5.02$ TeV (solid lines), for three different centrality bins,  $0$-$5$\% [panel (a)], $30$-$40$\% [panel (b)], and $60$-$70$\% [panel (c)]. The dashed lines represent the correlation coefficients with the correction for multiplicity fluctuations [Eq. (\ref{eq:multcorrect})]. The horizontal lines represent the correlation coefficients $\rho([p_T],v_2^2)$ between the mean transverse momentum and the momentum averaged harmonic flow.  }
	\label{fig:ptv2qv2q}
\end{figure}

The azimuthal anisotropy  of the momentum distribution of the hadrons emitted in a
heavy-ion collision can be described by the Fourier expansion,
\begin{equation}
\frac{d^2N}{dp d\phi} = \frac{dN}{2\pi dp} \left(  1 + 2 \sum_{n=1}^{\infty} V_n(p)e^{i n \phi} \right) \ \ ,
\label{ptdist}
\end{equation}
where 
$V_n(p)=v_n(p) e^{i\Psi_n(p)}$ is the flow vector for the $n$th order harmonic flow.
The mean transverse momentum of particles emitted in an event is defined as,
\begin{equation}
[p_T]=\frac {1}{N} \int_{p_{min}}^{p_{max}}   dp \ p \frac{dN}{dp}
\end{equation}
and the momentum averaged harmonic flow coefficient is 
\begin{equation}
V_n =  \frac{1}{N}  \int_{p_{min}}^{p_{max}}  dp  V_n(p) \frac{dN}{dp} \ ,
\end{equation}
where $N$ is the multiplicity in the event, given by
\begin{equation}
N=\int_{p_{min}}^{p_{max}} dp   \frac{dN}{dp} \ .
\end{equation}

The correlation coefficient between the harmonic flow and the mean transverse momentum is defined as \cite{Bozek:2016yoj}
\begin{equation}
  \rho([p_T],v_n^2)= \frac{Cov([p_t],v_n^2)}{\sqrt{Var([p_T])Var(v_n^2)}} \ ,
  \label{eq:ptv}
  \end{equation}
where the covariance,
\begin{equation}
  Cov([p_T],v_n^2)=\langle [p_T] V_n V_n^\star \rangle -\langle[p_t] \rangle \langle V_n V_n^\star\rangle \ ,
  \end{equation}
and the variances,
\begin{equation}
  Var([p_t])=\langle [p_T]^2 \rangle -\langle[p_t] \rangle^2,
\end{equation}
\begin{equation}
  Var(v_n^2)=\langle \left(V_n V_n^\star \right)^2\rangle -\langle V_n V_n^\star \rangle^2 \ ,
\end{equation}
are obtained as averages, $\langle \dots \rangle$, over the events in a given centrality bin, with the selfcorrelations in the sum over particles in an event excluded. In this paper, we use the boost invariant viscous hydrodynamic model MUSIC \cite{Schenke:2010nt,Schenke:2010rr,Paquet:2015lta}, with the Glauber \cite{Bozek:2019wyr} or the TRENTO model \cite{Moreland:2014oya} for the initial conditions for Pb+Pb and U+U collisions.  The details of the calculation and the parameters used can be found in Refs. \cite{Bozek:2021mov,Samanta:2023qem}. 

\begin{figure}
	\vspace{5mm}
	\begin{center}
	  \includegraphics[width=0.4\textwidth]{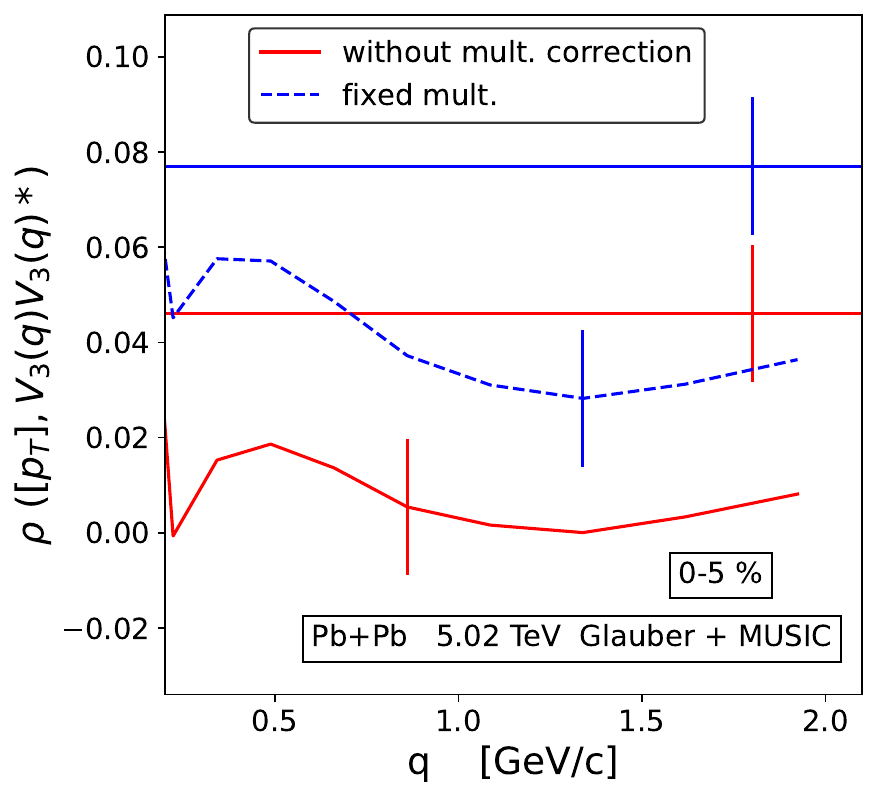} 
 	\end{center}
	\caption{The momentum dependent correlation coefficients $\rho([p_T],V_3(q)V_3(q)^\star)$ in Pb+Pb collisions at $\sqrt{s_{NN}}=5.02$ TeV for $0$-$5$\% centrality. The legends are similar to Fig. \ref{fig:ptv2qv2q}. }
	\label{fig:ptv3qv3q}
\end{figure}

\begin{figure}
	\vspace{5mm}
	\begin{center}
	  \includegraphics[width=0.4\textwidth]{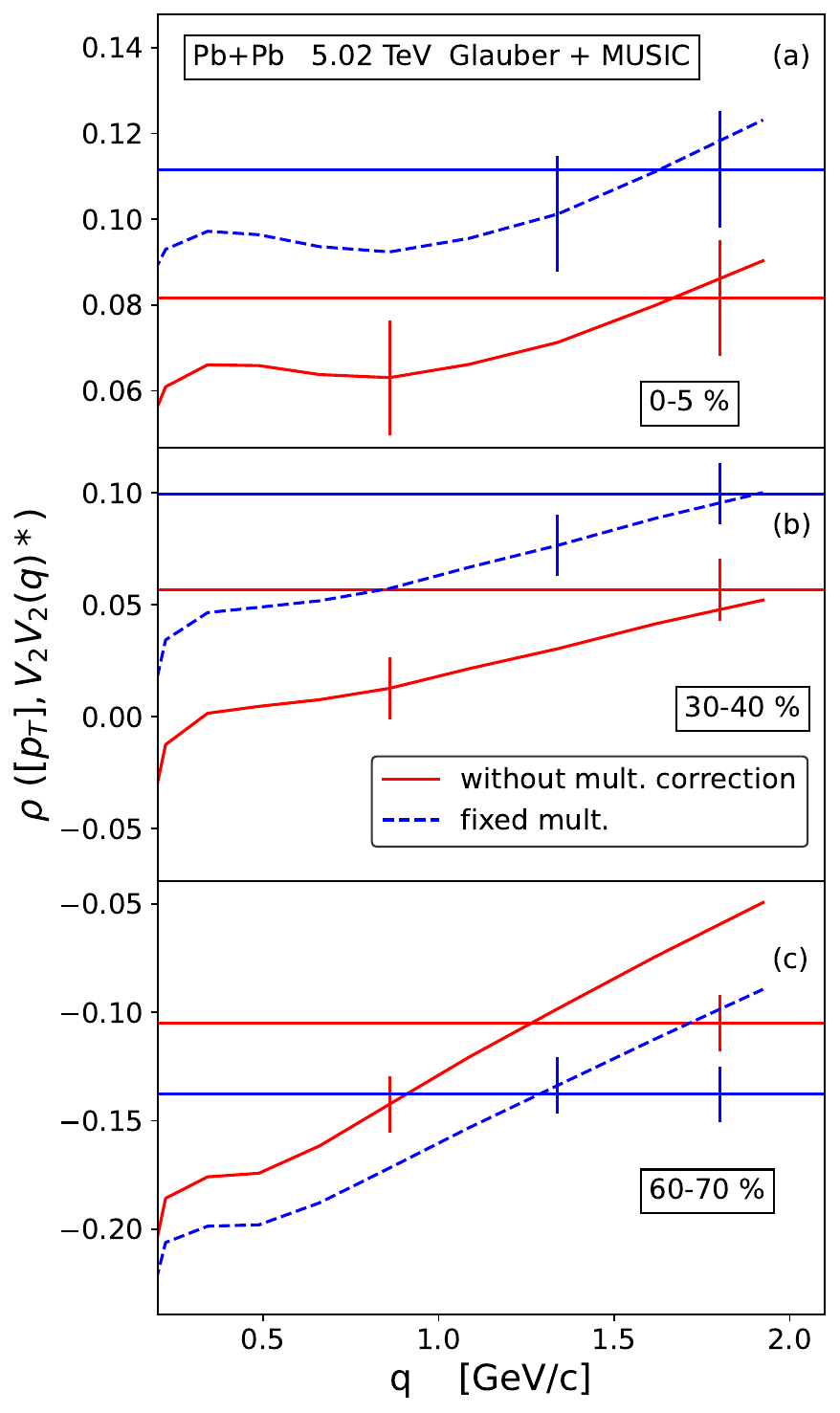} 
	\end{center}
	\caption{Same as in Fig. \ref{fig:ptv2qv2q}, but for the correlation coefficient $\rho([p_T],V_2V_2(q)^\star)$.}
	\label{fig:ptv2v2q}
\end{figure}

\begin{figure}
	\vspace{5mm}
	\begin{center}
	  \includegraphics[width=0.4\textwidth]{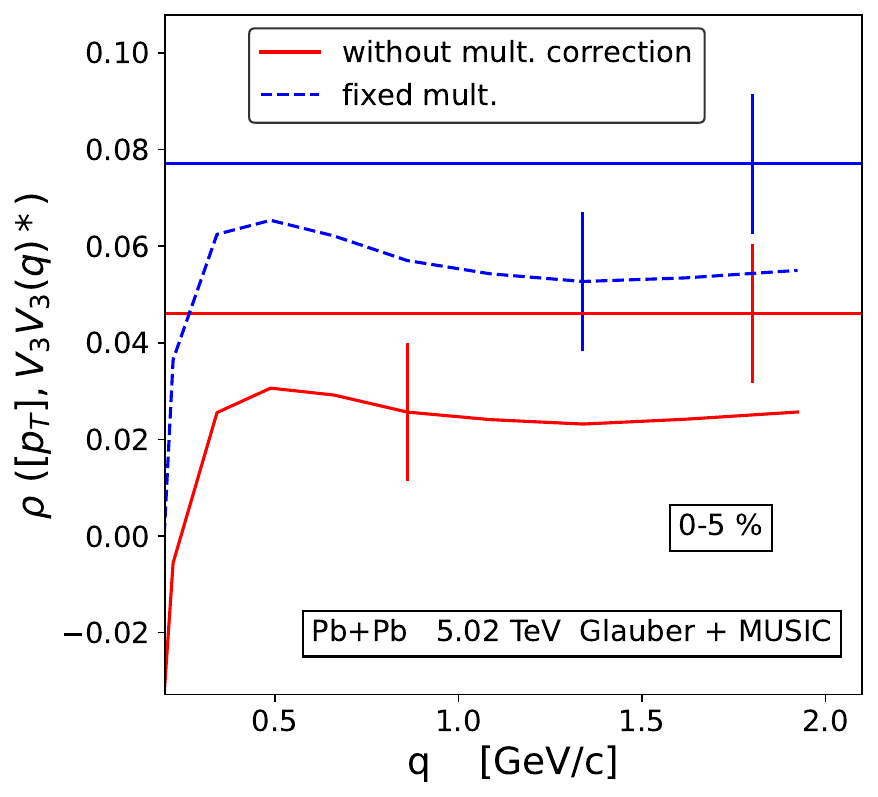}
 	\end{center}
	\caption{Same as in Fig. \ref{fig:ptv3qv3q}, but for the correlation coefficient $\rho([p_T],V_3V_3(q)^\star)$.}
	\label{fig:ptv3v3q}
\end{figure}

The momentum dependent correlation coefficient can be constructed as the correlation coefficient  between the mean
transverse momentum and the harmonic flow in a given transverse momentum bin,
\begin{equation}
  \rho\left([p_T],V_n(q)V_n(q)^\star\right) = \frac{Cov\left([p_T],V_n(q)V_n(q)^\star\right)}{\sqrt{Var\left([p_T]\right)Var\left(V_n(q)V_n(q)^\star\right)}} \ .
  \label{eq:ptvqq}
  \end{equation}
The correlation coefficient is a function of the transverse momentum $q$; it  should not be confused with $[p_T]$, which is not a variable. Here, we use the notation $V_n(q)V_n(q)^\star$ for the momentum dependent harmonic flow instead of $v_n(q)^2$ in order to distinguish it easily from the quantity $V_nV_n(q)^\star$ that we  discuss latter.
The results for the correlation coefficient in Eq. (\ref{eq:ptvqq}) for the elliptic and the triangular flow in Pb+Pb collisions, obtained from event-by-event viscous hydrodynamic simulations with Glauber initial conditions, are shown in Figs. \ref{fig:ptv2qv2q} and \ref{fig:ptv3qv3q} for $q<2$ GeV.
Note that  measurements at higher $q$ could be interesting for the study of
non-flow effects or  correlations originating from the color-glass condensate. For clarity, in the figures we show
the statistical errors of the simulation only for some points on the plot.

We notice a strong dependence of the  correlation coefficients on the transverse  momentum $q$ for the elliptic flow. This dependence explains the experimentally observed dependence of the momentum independent correlation coefficient, $\rho([p_T],v_n^2)$, on the transverse momentum cuts
\cite{ATLAS:2019pvn}. The  momentum dependent  coefficient, $\rho\left([p_T],V_n(q)V_n(q)^\star\right)$, is a measure of the correlation between the mean transverse momentum and the amount of harmonic flow at a definite transverse momentum $q$, irrespective of the specific $q$ dependence of the harmonic flow, $\langle v_n(q)^2\rangle$.
\begin{figure}
	\vspace{5mm}
	\begin{center}
	  \includegraphics[width=0.4\textwidth]{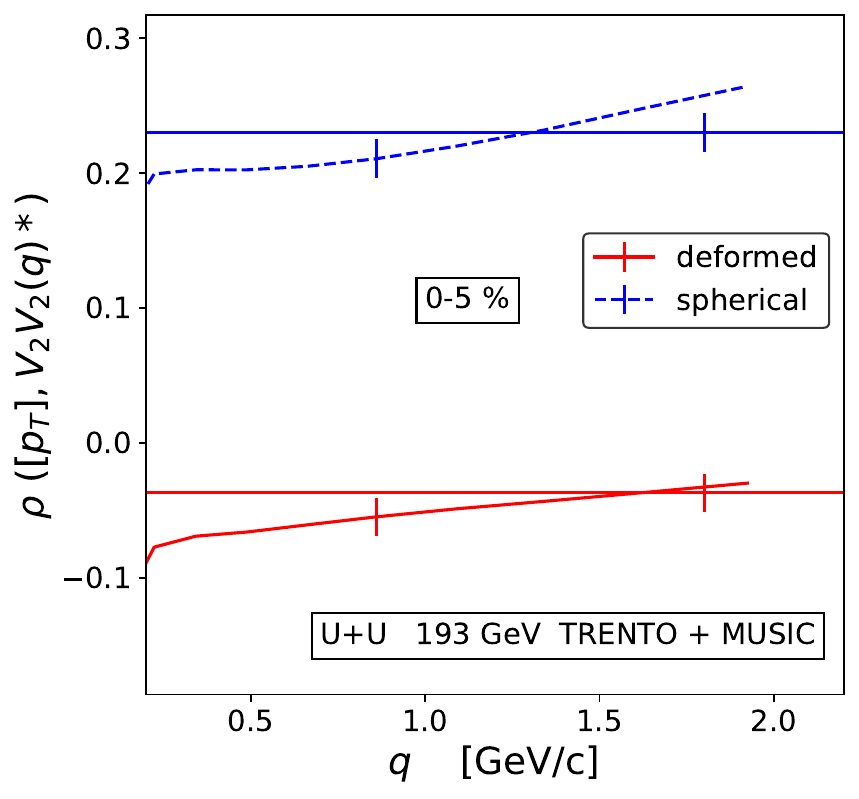} \\
 	\end{center}
	\caption{The momentum dependent correlation coefficients $\rho([p_T],V_2V_2(q)^\star)$  for U+U collisions at $\sqrt{s_{NN}}=193$GeV for $0$-$5$\% centrality. The solid line is for collisions of spherical nuclei, while the dashed line represents the results for deformed nuclei.}
	\label{fig:ptv2v2UU}
\end{figure}
The momentum independent correlation coefficients, $\rho([p_T],v_n^2)$ are plotted in  Figs. \ref{fig:ptv2qv2q} and \ref{fig:ptv3qv3q}, as horizontal solid lines. Please note that the correlation coefficient for the momentum averaged flow, $\rho([p_T],v_n^2)$, is not a momentum average of the momentum dependent correlation coefficient, $\rho\left([p_T],V_n(q)V_n(q)^\star\right)$. We have checked that it is due to the construction of the correlation coefficient as a ratio of two momentum dependent averages. In Sec. \ref{sec:other}, we compare the two covariances, $Cov\left([p_T],V_n(q)V_n(q)^\star\right)$ and $Cov\left([p_T],v_n^2\right)$ directly.

The correlation coefficient between the mean transverse momentum and the harmonic flow depends on the way the centrality bins are defined. In our calculation, we use wide bins with  the total entropy deposited in the collision as the
quantity determining the centrality. Large multiplicity fluctuations in each
centrality bin influence the correlation coefficient. This can be corrected by removing the multiplicity fluctuation from the statistical measures used \cite{Olszewski:2017vyg,Schenke:2020uqq}.
Each quantity $\mathcal{O}$ estimated in a given event is corrected as follows
\begin{equation}
  \mathcal{O}_{corr}=\mathcal{O}-\frac{Cov\left(\mathcal{O},N\right)}{\sqrt{Var\left(\mathcal{O}\right)Var\left(N\right)}}\left(N-\langle N \rangle\right) \ .
  \label{eq:multcorrect}
\end{equation}
The correlation coefficients for the mean transverse momentum and the harmonic flow coefficients corrected for multiplicity fluctuations are shown with dashed lines in Figs. \ref{fig:ptv2qv2q} and \ref{fig:ptv3qv3q}.
Corrections for multiplicity fluctuations are numerically sizable and should be used depending on the centrality definition used in the experiment.  Unless stated otherwise, we use quantities corrected for multiplicity  fluctuations in the following.

The momentum dependent correlation coefficient between the mean transverse momentum and the harmonic flow can also be defined as,
\begin{equation}
  \rho\left([p_T],V_nV_n(q)^\star\right) = \frac{Cov\left([p_T],V_nV_n(q)^\star\right)}{\sqrt{Var\left([p_T]\right)Var\left(V_nV_n(q)^\star\right)}} \ .
  \label{eq:ptvq}
  \end{equation}
The above correlation coefficient could be easier to  measure experimentally. In the denominator, $Var\left( V_n V_n(q)^\star\right)$, unlike  $Var\left( V_n(q) V_n(q)^\star\right)$, is a four particle correlator with only two particles restricted to a limited transverse momentum bin $q$. However, the correlation coefficient (\ref{eq:ptvq}) does not have such a simple interpretation like the coefficient (\ref{eq:ptvqq}). The results for $\rho\left([p_T],V_nV_n(q)^\star\right)$, from hydrodynamic simulations, are presented in Figs. \ref{fig:ptv2v2q} and \ref{fig:ptv3v3q}. 
The qualitative behavior on the transverse momentum  ($q<2$GeV) remains similar for the correlation coefficients  $\rho\left([p_T],V_nV_n(q)^\star\right)$ and $\rho\left([p_T],V_n(q)V_n(q)^\star\right)$. However, 
the dependence of the correlation coefficients,  $\rho\left([p_T],V_nV_n(q)^\star\right)$, on the transverse momentum $q$ is weaker than for the correlation coefficient,  $\rho\left([p_T],V_n(q)V_n(q)^\star\right)$.

\begin{figure}
	\vspace{5mm}
	\begin{center}
	  \includegraphics[width=0.4\textwidth]{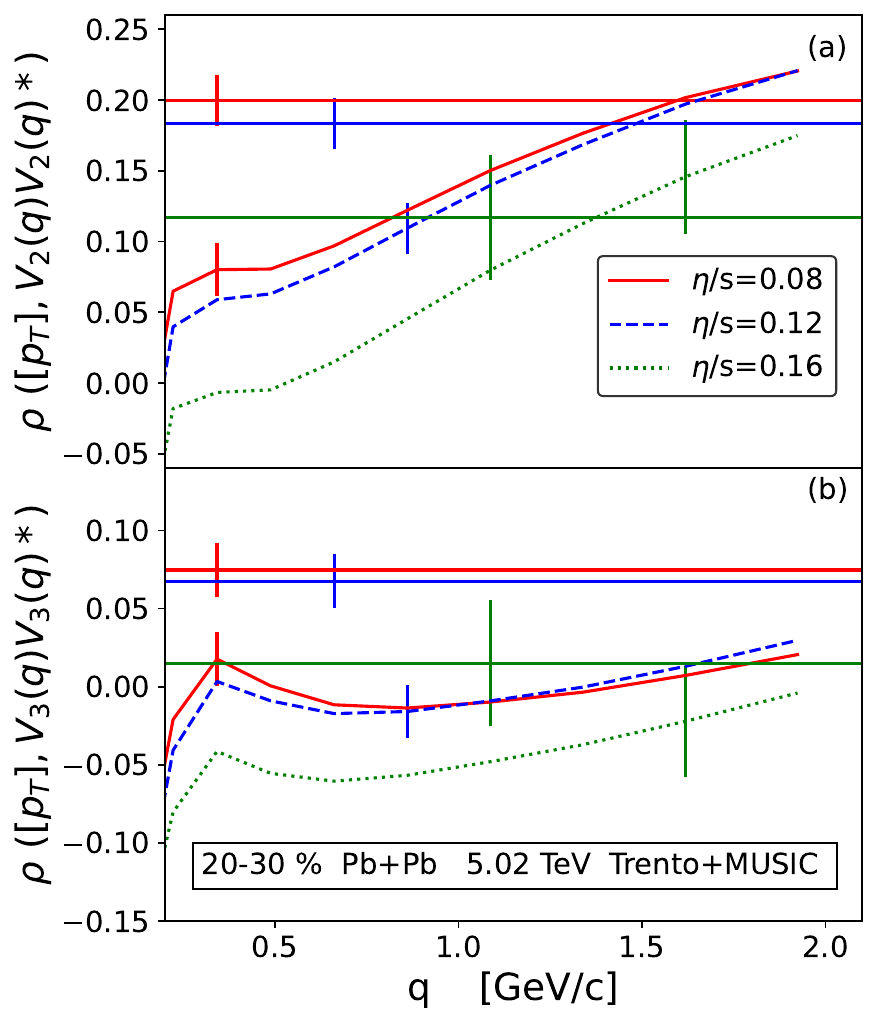} \\
 	\end{center}
	\caption{The momentum dependent correlation coefficients $\rho([p_T],V_nV_n(q)^\star)$  for Pb+Pb collisions at $\sqrt{s_{NN}}=5.02$TeV for $20$-$30$\% centrality, for the elliptic flow [panel (a)] and the triangular flow [panel (b)]. Results for three different values of the shear viscosity to entropy density are shown $\eta/s=0.08,$ $0.12$, and $0.16$ using solid, dashed and dotted lines respectively. }
	\label{fig:rhoshear}
\end{figure}

In collisions of deformed nuclei averaging over the relative orientation of the two colliding nuclei reduces the value of the momentum independent correlation coefficient \cite{Giacalone:2019pca}. This effect is visible in Fig. \ref{fig:ptv2v2UU} as the difference between the horizontal lines representing the results for 
the momentum independent correlation coefficient in collisions of spherical and deformed nuclei. The corresponding momentum dependent coefficients are shown with the dashed and solid curves respectively. The momentum dependence is qualitatively similar for the two cases, for $q<2$GeV.  It indicates that the momentum dependent correlation coefficient is not specifically sensitive to the global shape fluctuations coming from the nuclear deformation.

In Fig. \ref{fig:rhoshear} are presented the momentum dependent correlation coefficients
for three different values of the shear viscosity to entropy density ratio $\eta/s$. Qualitatively the dependence on shear viscosity is similar to the dependence on nuclear deformation. The change in shear viscosity causes a shift of the curves, without modifying strongly the shape of their  momentum dependence. The shift is much smaller in magnitude than for the nuclear deformation shown in Fig. \ref{fig:ptv2v2UU}. The value of the shear viscosity  influences the correlation coefficient, causing a shift of the curves. However,  the momentum dependent version of the correlation coefficients is not specifically sensitive to shear viscosity as compared to the momentum average one (solid lines in Fig. \ref{fig:rhoshear}), the shape of the curves is similar for different values of shear viscosity. We have checked that the dependence on  bulk viscosity of the considered correlation coefficients is similar to the dependence on shear viscosity.

\section{Granularity in the initial state}

\label{sec:granularity}

\begin{figure}
	\vspace{5mm}
	\begin{center}
		\includegraphics[width=0.4\textwidth]{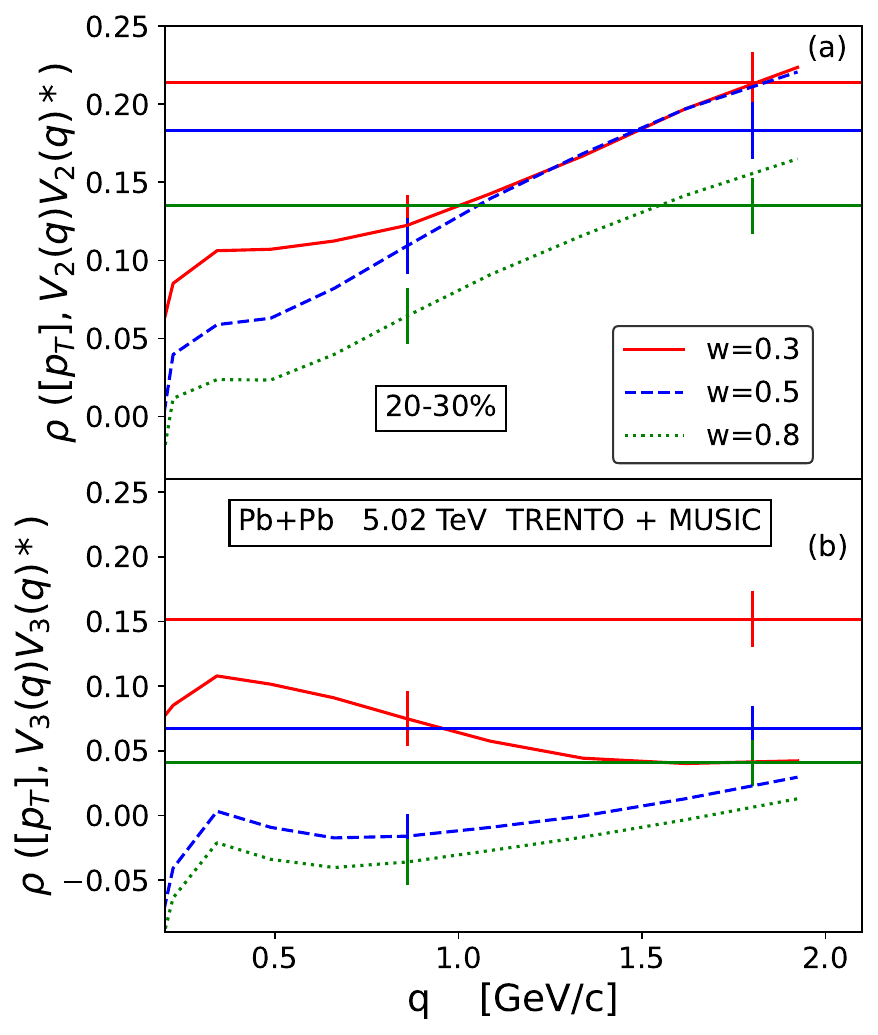}
	\end{center}
	\caption{The momentum dependent correlation coefficients $\rho([p_T],V_2(q)V_2(q)^\star)$ [panel (a)] and  $\rho([p_T],V_3(q)V_3(q)^\star)$ [panel (b)] in Pb+Pb collisions at $\sqrt{s_{NN}}=5.02$ TeV  for $20$-$30$\% centrality. Three different values of the width of the nucleons used for the energy deposition in the initial state: $w=0.3$, $0.5$, and $0.8$ fm are denoted with solid, dashed, and dotted lines respectively. The horizontal solid lines with corresponding color represent the momentum independent correlations coefficients $\rho([p_T],v_n^2)$.}
	\label{fig:ptvqvqw}
\end{figure}

The correlation coefficient between the transverse momentum and the harmonic flow is  sensitive to the granularity of the initial state for the hydrodynamic evolution 
\cite{Bozek:2016yoj,Giacalone:2021clp}.
 In models, the granularity of the initial state can be modified by changing the size of the region, where each of the participant nucleons deposits the initial energy. Experimental results suggest that the size of that region is small, which corresponds to an initial state with high granularity \cite{Giacalone:2021clp}. We study this size dependence by changing the two-dimensional Gaussian width associated with each nucleon in the TRENTO  model, with  $w=0.3$, $0.5$, and $0.8$ fm  \cite{Moreland:2014oya}.

\begin{figure}
	\vspace{5mm}
	\begin{center}
	  \includegraphics[width=0.4\textwidth]{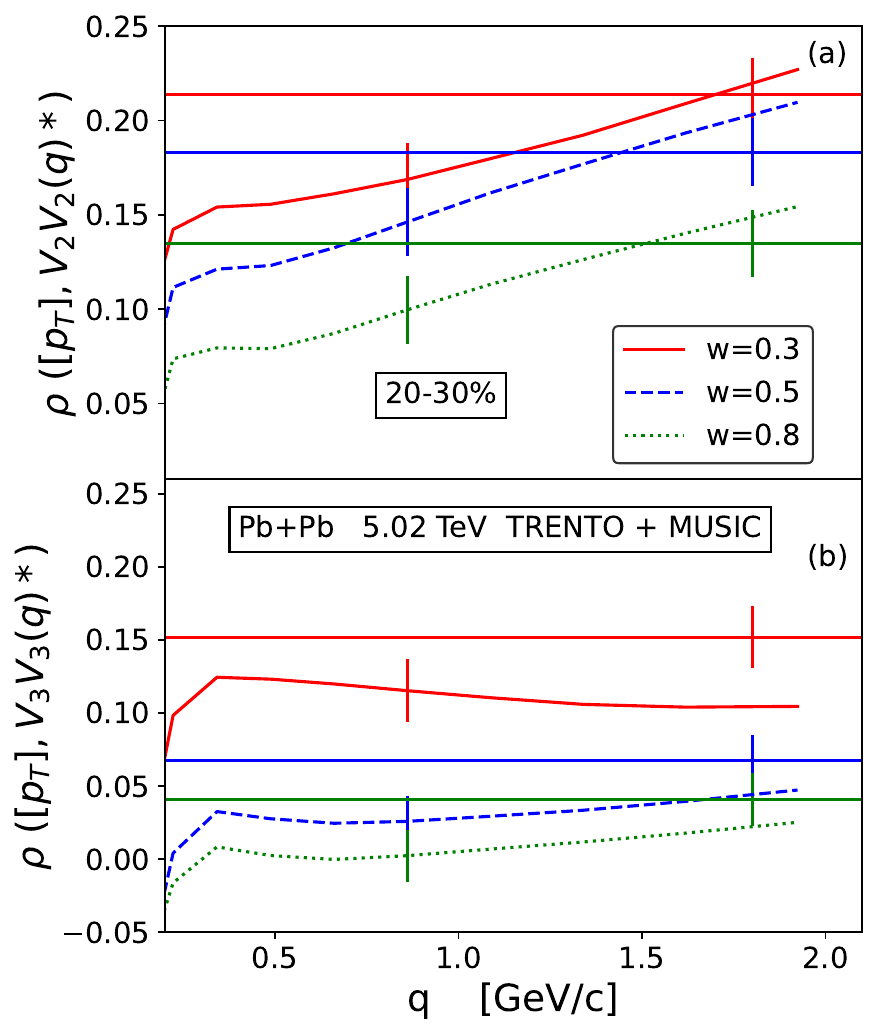}
 	\end{center}
	\caption{The momentum dependent correlation coefficients $\rho([p_T],V_2V_2(q)^\star)$ [panel (a)] and  $\rho([p_T],V_3V_3(q)^\star)$ [panel (b)] in Pb+Pb collisions at $\sqrt{s_{NN}}=5.02$ TeV  for $20$-$30$\% centrality. Three different values of the width of the nucleons used for the energy deposition in the initial state: $w=0.3$, $0.5$, and $0.8$ fm are denoted with solid, dashed, and dotted lines respectively. The horizontal solid lines with corresponding color represent the momentum independent correlations coefficients $\rho([p_T],v_n^2)$.} 
	\label{fig:ptvvqw}
\end{figure}

The momentum dependent correlation coefficients,  $\rho([p_T],V_n(q)V_n(q)^\star)$, for $20$-$30$\% centrality
are shown in Fig. \ref{fig:ptvqvqw}. The  correlation coefficients show a strong dependence on the transverse momentum $q$. The increase of the correlation coefficient with $q$ is less steep for a more granular initial state (smaller $w$). This effect is  stronger for the triangular flow  [Fig. \ref{fig:ptvqvqw} (b)], and the correlation even decreases with q for $w=0.3$ fm. In particular, the correlation coefficients show different momentum dependence for $q=0$-$1.5$ GeV for different values of $w$. The momentum independent correlation coefficients (baselines) follow a particular dependence on the granularity; it decreases as the granularity decreases ($w$ increases).   

The correlation coefficients,  $\rho([p_T],V_nV_n(q)^\star)$, show a quite similar dependence on the transverse momentum  in Fig. \ref{fig:ptvvqw}. Again, for the triangular flow the dependence on $q$ is less steep for the initial state with higher granularity and the difference between different initial states is the strongest in the range $q=0$-$1.5$ GeV. It would be interesting to compare model predictions and  experimental results not only for the momentum independent correlation coefficients \cite{Giacalone:2021clp}, but also for the momentum dependent correlation coefficients,  $\rho([p_T],V_n(q)V_n(q)^\star)$ or  $\rho([p_T],V_nV_n(q)^\star)$, in order to constrain the parameters of the initial state in the  hydrodynamic modeling of heavy-ion collisions. Results in   Fig. \ref{fig:ptvvqw}  indicate that the momentum dependent correlation coefficient $\rho([p_T],V_nV_n(q)^\star)$ serves as a good candidate to probe the granularity, while being easier  to measure  in experiments than  $\rho([p_T],V_n(q)V_n(q)^\star)$. The effect of granularity is further discussed in Sec. \ref{sec:covariance} through the momentum dependent covariances.

\section{Other momentum dependent measures of correlations between $[p_T]$ and $V_n(q)$}

\label{sec:other}

\begin{figure}
	\vspace{5mm}
	\begin{center}
	  \includegraphics[width=0.4\textwidth]{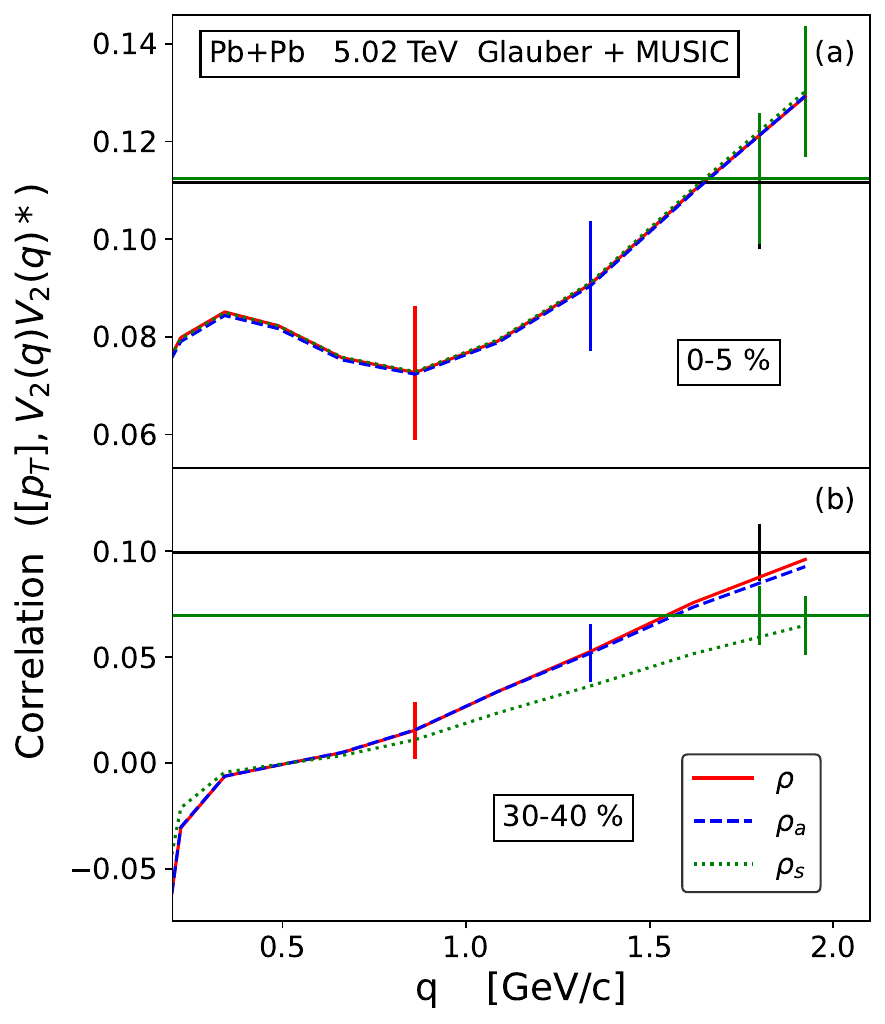}
	\end{center}
	\caption{The momentum dependent correlation coefficient $\rho\left([p_T],V_2(q)V_2(q)^\star\right)$ (solid lines), the approximate correlation coefficient $\rho_a\left([p_T],V_2(q)V_2(q)^\star\right)$ (dashed lines), and the scaled correlation coefficient $\rho_s\left([p_T],V_2(q)V_2(q)^\star\right)$ (dotted lines).
        The results are obtained  for Pb+Pb collisions at $\sqrt{s_{NN}}=5.02$ TeV in two different centrality bins, $0$-$5$\% [panel (a)] and $30$-$40$\% [panel (b)]. }
	\label{fig:measqq2}
\end{figure}

The experimental estimation of the momentum dependent correlation coefficients, $\rho([p_T],V_n(q)V_n(q)^\star)$ or  $\rho([p_T],V_nV_n(q)^\star)$, is more difficult than for the momentum independent correlation coefficient, 
$\rho([p_T],V_nV_n^\star)$. In particular, the estimate of the  variances,
$Var\left( V_n(q)V_n(q)^\star \right)$ or $Var\left( V_nV_n(q)^\star \right)$, requires the measurement of a
four particle correlator in a restricted transverse momentum bin. We check, if alternative, approximate formulas for the momentum dependent correlation coefficients can be used instead.

One possibility is to use the momentum averaged variance, $Var\left( v_n^2 \right)$, in the denominator, since it is easier to estimate a four particle correlator in the full acceptance.
The properly rescaled formulas for such correlation coefficients are, 
\begin{eqnarray}
  && \rho_a\left([p_T],V_n(q)V_n(q)^\star\right) = \nonumber \\
  &&\frac{Cov\left([p_T],V_n(q)V_n(q)^\star\right) \langle v_n^2\rangle}{\sqrt{Var\left([p_T]\right)Var\left( v_n^2\right)} \langle V_n(q)V_n(q)^\star \rangle }
  \label{eq:qqapprox}
\end{eqnarray}
and
\begin{eqnarray}
  && \rho_a\left([p_T],V_nV_n(q)^\star\right) = \nonumber \\
  &&\frac{Cov\left([p_T],V_nV_n(q)^\star\right) \langle v_n^2\rangle}{\sqrt{Var\left([p_T]\right)Var\left( v_n^2\right)} \langle V_n V_n(q)^\star \rangle } \ .
  \label{eq:qapprox}
\end{eqnarray}
The approximate formulas (\ref{eq:qqapprox}) and (\ref{eq:qapprox}) are expected to reproduce closely the original momentum dependent correlation coefficient, because the factors
\begin{equation}
  \frac{ \sqrt{Var\left(V_n(q)V_n(q)^\star\right)}\langle v_n^2\rangle}{\sqrt{Var\left( v_n^2\right)} \langle V_n(q) V_n(q)^\star \rangle}
\end{equation}
and
\begin{equation}
  \frac{\sqrt{Var\left( V_nV_n(q)^\star\right)} \langle v_n^2\rangle}{\sqrt{Var\left( v_n^2\right)} \langle V_n V_n(q)^\star \rangle}
\end{equation}
are
very close to $1$ \cite{Bozek:2021mov}.

\begin{figure}
	\vspace{5mm}
	\begin{center}
	  \includegraphics[width=0.4\textwidth]{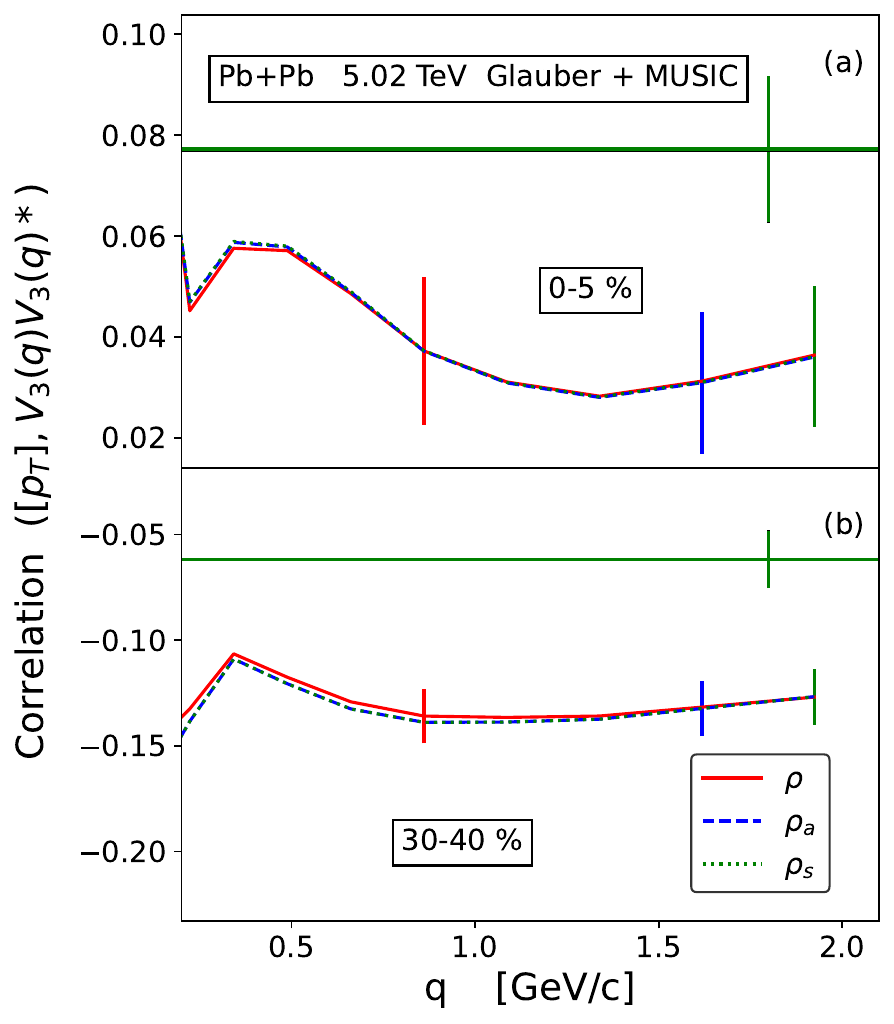}
	\end{center}
	\caption{Same as in Fig. \ref{fig:measqq2} but for the triangular flow.}
	\label{fig:measqq3}
\end{figure}

Another possible construction would be to scale the covariance between the mean transverse momentum and the harmonic flow by the mean of the harmonic flow squared instead of  its standard deviation. The formulas for the scaled correlation coefficients are,
\begin{equation}
   \rho_s\left([p_T],V_n(q)V_n(q)^\star\right) = 
   \frac{Cov\left([p_T],V_n(q)V_n(q)^\star\right) }{\sqrt{Var\left([p_T]\right)}\langle V_n(q)V_n(q)^\star\rangle)} 
   \label{eq:qqscal}
  \end{equation}
and
\begin{equation}
   \rho_s\left([p_T],V_nV_n(q)^\star\right) = 
   \frac{Cov\left([p_T],V_nV_n(q)^\star\right) }{\sqrt{Var\left([p_T]\right)} \langle V_nV_n(q)^\star\rangle }
    \ .
   \label{eq:qscal}
\end{equation}
The scaled correlation coefficient, $\rho_s$, is expected to be a good approximation of the original correlation $\rho$ for a fluctuation dominated harmonic flow, when $ \sqrt{Var\left(V_n(q)V_n(q)^\star\right)} \simeq \langle V_n(q) V_n(q)^\star \rangle$  and  $ \sqrt{Var\left(V_nV_n(q)^\star\right)} \simeq \langle V_n V_n(q)^\star \rangle$, e.g. elliptic flow in central collision and triangular flow in general. The momentum independent version of the scaled correlation coefficient,
\begin{equation}
   \rho_s\left([p_T],v_2^2\right) = 
   \frac{Cov\left([p_T],v_2^2\right) }{\sqrt{Var\left([p_T]\right)} \langle v_2^2\rangle }
    \ ,
    \label{eq:qscalbase}
\end{equation}
is  the normalized symmetric cumulant between the mean transverse momentum $[p_T]$ and the harmonic flow $v_n^2$ \cite{Bozek:2021zim}. It 
has been used by the STAR Collaboration  to analyze the nuclear deformation in relativistic Au+Au and U+U collisions \cite{JJcph}.

\begin{figure}
	\vspace{5mm}
	\begin{center}
	  \includegraphics[width=0.4\textwidth]{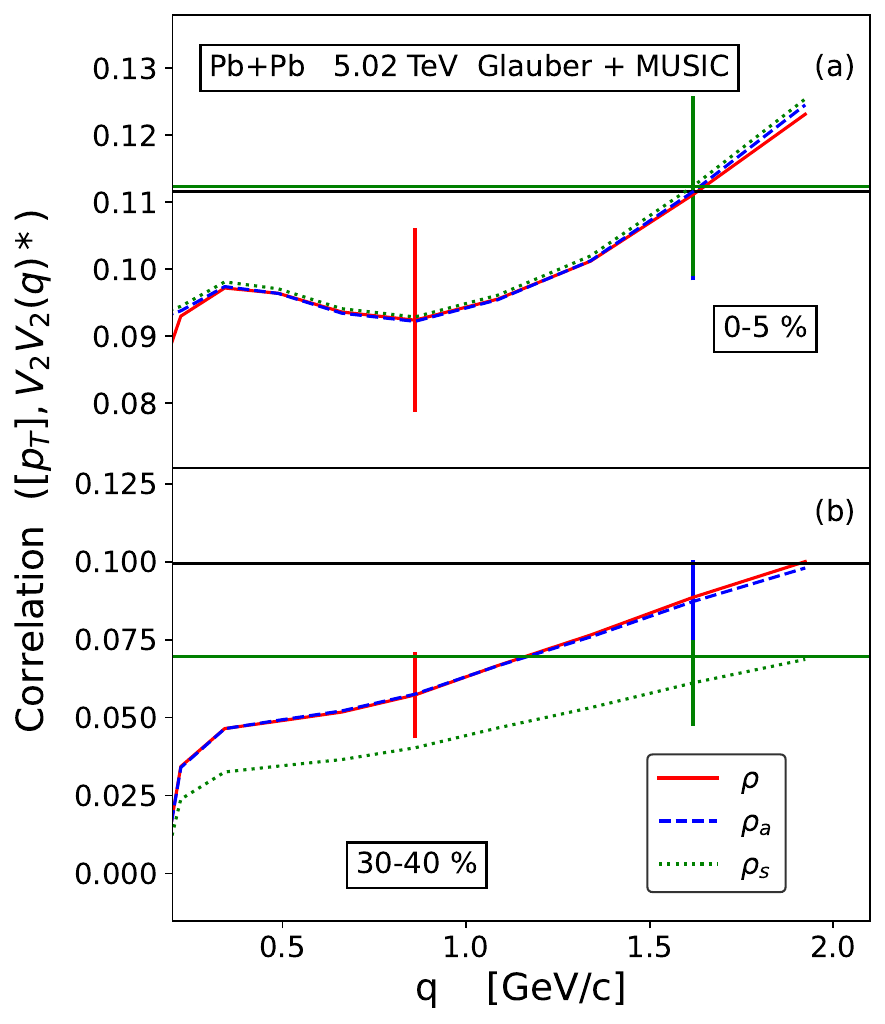}
	\end{center}
	\caption{The momentum dependent correlation coefficient $\rho\left([p_T],V_2V_2(q)^\star\right)$ (solid lines), the approximate correlation coefficient $\rho_a\left([p_T],V_2V_2(q)^\star\right)$ (dashed lines), and the scaled correlation coefficient $\rho_s\left([p_T],V_2V_2(q)^\star\right)$ (dotted lines).
		The results are obtained  for Pb+Pb collisions at $\sqrt{s_{NN}}=5.02$ TeV in two different centrality bins, $0$-$5$\% [panel (a)] and $30$-$40$\% [panel (b)]. }
	\label{fig:measq2}
\end{figure}

\begin{figure}
	\vspace{5mm}
	\begin{center}
	  \includegraphics[width=0.4\textwidth]{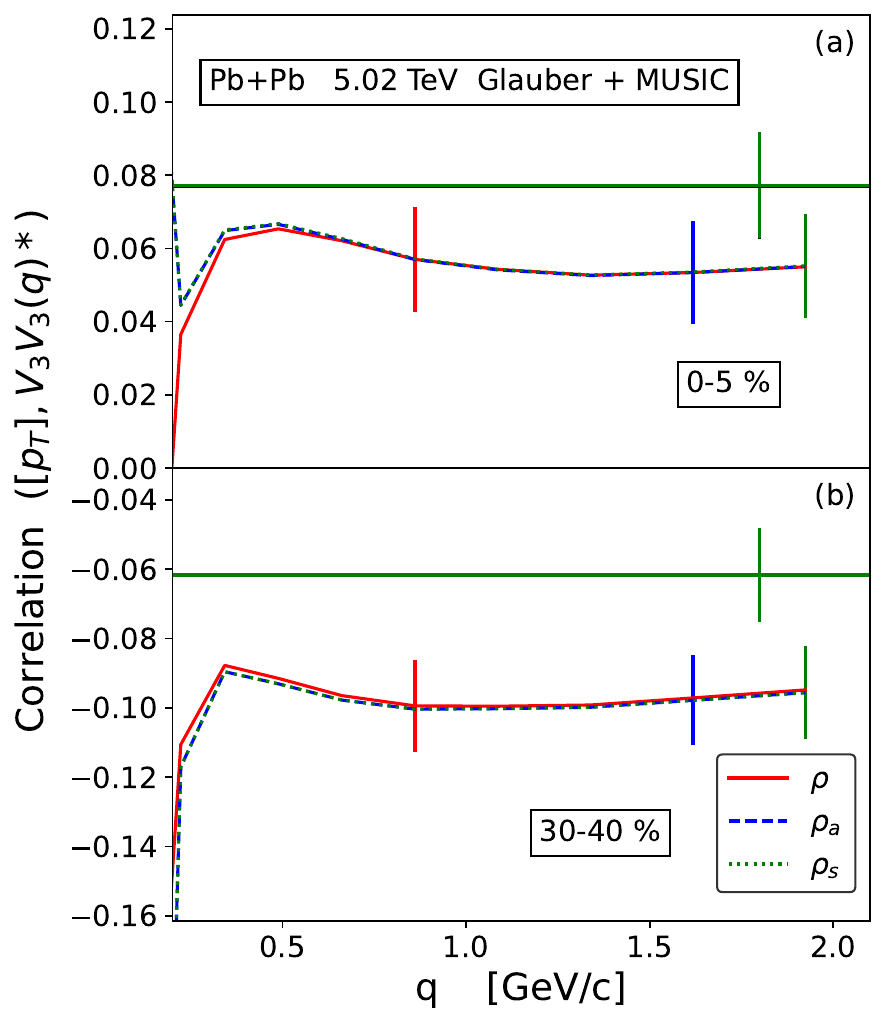}
	\end{center}
	\caption{Same as in Fig. \ref{fig:measq2} but for the triangular flow. }
	\label{fig:measq3}
\end{figure}

The comparison between the full correlation coefficient and the approximate expressions are  presented in Figs. \ref{fig:measqq2} and \ref{fig:measqq3} for the momentum dependent correlation coefficient $\rho\left([p_T],V_n(q)V_n(q)^\star\right)$ and in Figs. \ref{fig:measq2} and \ref{fig:measq3} for the coefficient
$\rho\left([p_T],V_nV_n(q)^\star\right)$. In all cases and for $q<2$ GeV, the approximate correlation coefficient $\rho_a$ is close to the correlation coefficient $\rho$, and could be used as an experimental estimate thereof. The scaled correlation coefficient $\rho_s$ is a good approximation of the momentum dependent correlation coefficient $\rho$  for the triangular flow and for the elliptic flow in central collisions, i.e. in cases with a fluctuation dominated harmonic flow. Please note that all of the momentum dependent  correlation
coefficients discussed in this section, $\rho_a$ or $\rho_s$, are well defined observables that could be  measured in experiment and compared to model calculations, even if they are not exactly the Pearson correlation coefficient of $[p_T]$ and the momentum dependent harmonic flow.

  \section{Scaled covariance}

  \label{sec:covariance}

\begin{figure}[th]
	\vspace{5mm}
	\begin{center}
	  \includegraphics[width=0.4\textwidth]{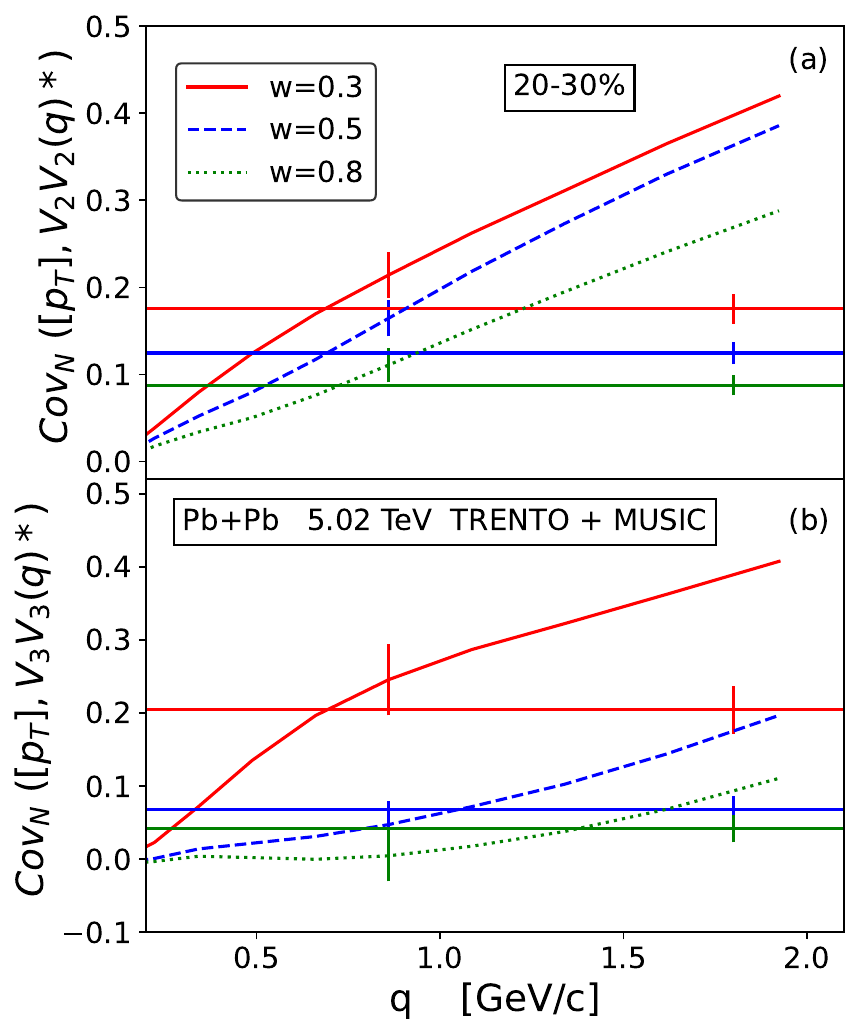}
	\end{center}
	\caption{The normalized covariance between the mean transverse momentum and the momentum dependent harmonic flow  [Eq. (\ref{eq:scaledcov})] in Pb+Pb collisions,  for the elliptic flow  [panel (a)] and for the triangular flow [panel (b)]. Three different nucleon widths used for the energy deposition in the initial state: $w=0.3$, $0.5$, and $0.8$ fm, are denoted with solid, dashed, and dotted lines respectively. The horizontal solid lines represent the momentum independent values for the scaled correlations coefficient $\rho_s([p_T],v_n^2)$ [Eq. (\ref{eq:qscalbase})]}
	\label{fig:NCov}
\end{figure}

\begin{figure}
	\vspace{5mm}
	\begin{center}
	  \includegraphics[width=0.4\textwidth]{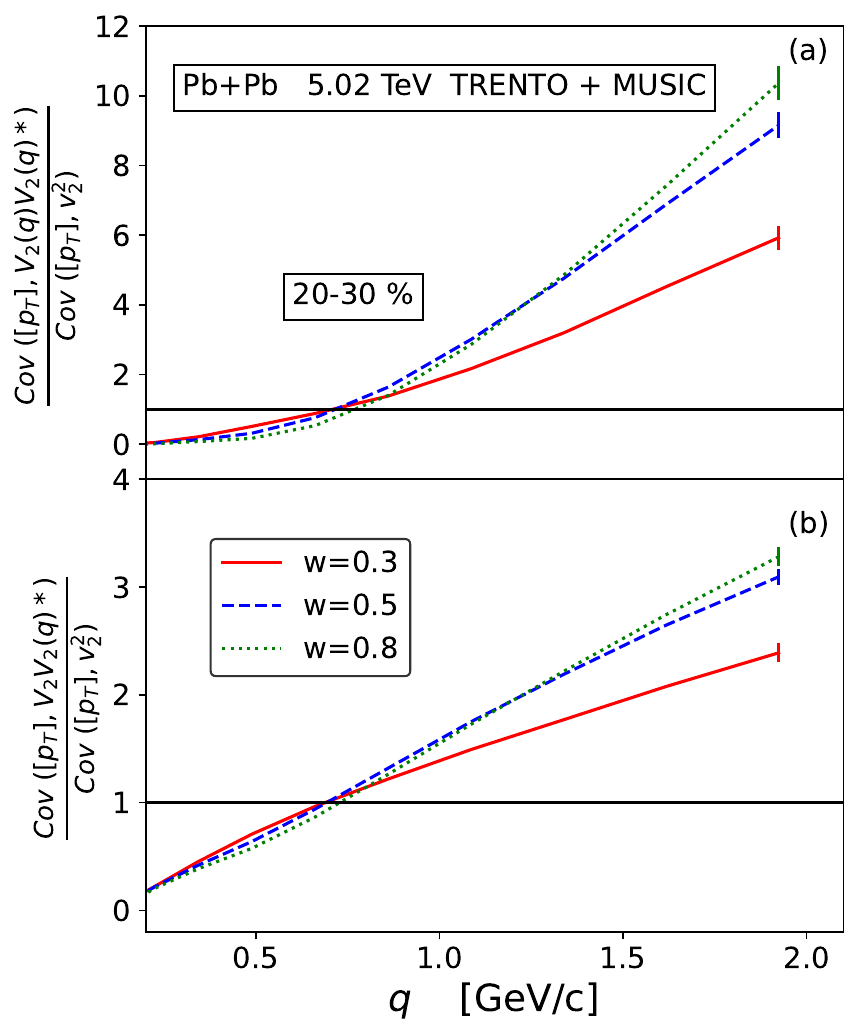} \\
	\end{center}
	\caption{The ratio of the momentum dependent and momentum independent covariance for the elliptic flow in Pb+Pb collisions, for $Cov\left([p_T],V_2(q)V_2(q)^\star\right)$ [panel (a)] and $Cov\left([p_T],V_2V_2(q)^\star\right)$ [panel (b)]. Three different values for the width of the nucleons used for the energy deposition in the initial state: $w=0.3$, $0.5$, and $0.8$ fm are denoted with solid, dashed, and dotted lines respectively}
	\label{fig:covarianceratio}
\end{figure}

  The momentum dependent correlation coefficients, $\rho\left([p_T],V_n(q)V_n(q)^\star\right)$ or  $\rho\left([p_T],V_nV_n(q)^\star\right)$, are defined as a ratio of two quantities that strongly depend on the transverse momentum $q$.
  It could be interesting
 to look directly at the momentum dependent covariances
  between the mean transverse momentum and the harmonic flow. In particular, a normalized covariance can be defined as,
  \begin{equation}
  Cov_N\left([p_T],V_n(q)V_n(q)^\star\right)=\frac{Cov\left([p_T],V_n(q)V_n(q)^\star\right) }{\sqrt{Var([p_T])} \langle v_n^2 \rangle}  \ .
  \label{eq:scaledcovqq}
  \end{equation}
  or 
\begin{equation}
  Cov_N\left([p_T],V_nV_n(q)^\star\right)=\frac{Cov\left([p_T],V_nV_n(q)^\star\right) }{\sqrt{Var([p_T])} \langle v_n^2 \rangle}  \ .
  \label{eq:scaledcov}
  \end{equation}
The normalization is chosen in a  way similar to that  for the normalized symmetric cumulant of the mean transverse momentum and the harmonic flow coefficients \cite{Bilandzic:2013kga,Mordasini:2019hut,Bozek:2021zim}. Also please note that the baseline for $Cov_N\left([p_T],V_n(q)V_n(q)^\star\right)$ and $Cov_N\left([p_T],V_nV_n(q)^\star\right)$ is given by $\rho_s\left([p_T],v_2^2\right)$ of Eq. (\ref{eq:qscalbase}).

\begin{figure}
	\vspace{5mm}
	\begin{center}
	  \includegraphics[width=0.4\textwidth]{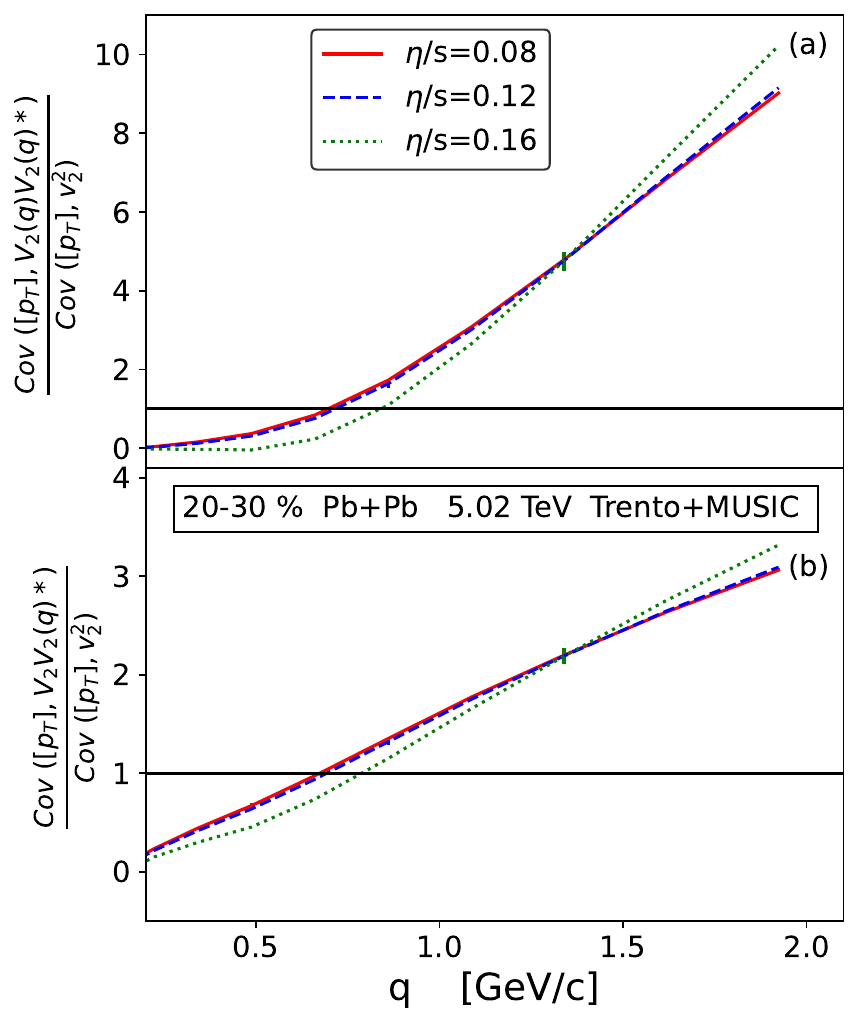} \\
	\end{center}
	\caption{The ratio of the momentum dependent and momentum independent covariance for the elliptic flow in Pb+Pb collisions, for $Cov\left([p_T],V_2(q)V_2(q)^\star\right)$ [panel (a)] and $Cov\left([p_T],V_2V_2(q)^\star\right)$ [panel (b)]. Three different values for the shear viscosity to entropy ratio $\eta/s=0.08$, $0.12$, and $0.16$ are denoted with solid, dashed, and dotted lines respectively.}
	\label{fig:covarianceratio_shear}
\end{figure}

The normalized covariances (\ref{eq:scaledcov}) for the elliptic and triangular flows in Pb+Pb collisions and for $20$-$30$\% centrality  are shown in Fig. \ref{fig:NCov}. It is  visible that the dependence on the transverse momentum $q$ of the normalized covariance between the mean transverse momentum and the harmonic flow is very sensitive to $w$, the size of the region of in the transverse plane, where each participant nucleon deposits the initial energy. The effect is particularly strong for the triangular flow, with the  steepest dependence for $w=0.3$ in the range of transverse momentum $q<1$ GeV making a striking difference from the other two cases. 

The momentum dependence of the covariance for different values of the parameter $w$, can also be directly compared by looking at the ratio
\begin{equation}
  \frac{Cov\left([p_T],V_n(q)V_n(q)^\star\right)}{Cov\left([p_T],V_nV_n^\star\right)} 
  \label{eq:covarianceratioqq}
\end{equation}
or
\begin{equation}
  \frac{Cov\left([p_T],V_nV_n(q)^\star\right)}{Cov\left([p_T],V_nV_n^\star\right)} 
  \label{eq:covarianceratio}
  \end{equation}
of the momentum dependent and momentum averaged covariances. This is possible whenever the denominator is not close to 0. The  simulation results for the covariance ratios in  Eqs. (\ref{eq:covarianceratioqq}) and  (\ref{eq:covarianceratio}) with three different values of the parameter $w$ are compared in  Fig. \ref{fig:covarianceratio}. The transverse momentum dependence of the covariance ratio is even more spectacular here. For both of the covariance ratios, all lines cross the baseline, $1$, at $q\simeq \langle [p_T] \rangle$ and then they split at higher momenta depending on the participant nucleon density deposition size $w$ producing a remarkable difference for $q \simeq 1$-$2$ GeV. This striking effect of the granularity on the covariance ratio could be studied in experiments to constrain the nucleon width with great precision.

In Fig. \ref{fig:covarianceratio_shear} is shown the dependence of the covariance ratios on the ratio of shear viscosity to entropy density $\eta/s$. We find that the momentum dependence of the covariance ratio shows a dependence on the value of shear viscosity. The shape of the momentum dependence could be used as an additional constraint on the value of shear viscosity in Bayesian analysis of model simulation and experimental data \cite{JETSCAPE:2020mzn,Nijs:2020roc}.

\section{Summary and Outlook}

We propose to measure momentum dependent correlation coefficients between the mean transverse momentum and the harmonic flow. The measurement of quantities based on the covariance between two observables, the mean transverse momentum and the harmonic flow in a given transverse momentum bin,  brings additional information on the  statistical multidimensional distribution of these flow observables. In the simplest version, it is
the correlation coefficient of the mean transverse momentum and the square of the harmonic flow vector in a given transverse momentum bin. A modified 
definition of such a correlation coefficient involves the first flow vector taken in the whole acceptance and the second flow vector in a given transverse momentum bin. Both
quantities would give new momentum dependent observables for correlations between the
transverse flow and harmonic flow variables.

The momentum dependent correlation coefficient  depends on physics parameters  used in the hydrodynamic modeling of the collision, e.g. the nuclear deformation, the shear viscosity, or the initial state granularity. We show that the form of the momentum dependence of the proposed correlation coefficients
is sensitive to the granularity of the initial state, especially, for the
triangular flow. Such a measurement could serve to additionally constrain  small scale fluctuations present in the initial state of nucleus-nucleus or proton-nucleus collisions. The proposed momentum dependent correlation coefficient or the momentum dependent covariance between the mean transverse momentum and the harmonic
flow can be be sensitive also to other physical parameters in the model, e.g. the value of shear and bulk viscosity. Therefore,  such  a complete analysis  could be done
using a Bayesian analysis combining experimental data on many observables, besides the correlations and covariances discussed in this paper,
and a set of model parameters \cite{JETSCAPE:2020mzn,Nijs:2020roc}.

We suggest that a simplified formula for the momentum dependent correlation coefficient could be used in experimental analyses  instead of the exact formula for the momentum dependent correlation coefficient. The simplified formula involves a four particle correlator in the denominator for particles in the whole acceptance and not in a limited transverse momentum bin. We note that a scaled or normalized covariance between the mean transverse momentum and the harmonic flow in a given transverse momentum bin is an interesting observable by itself, sensitive to the granularity in the initial state. Lastly, the ratio of the momentum dependent and the momentum averaged covariance shows a nice splitting and a very strong difference at  higher momenta according to the granularity. The experimental data on such simplified momentum dependent observables could be used to constrain parameters used in hydrodynamic models of heavy-ion collisions.

It would be very interesting to measure the momentum dependent correlation coefficients in relativistic collisions with heavy ions and/or protons. This could provide new information on the fluctuating initial state, the dynamics of the collision, the hadronization, or color glass correlations in such collisions.

\section*{Acknowledgments}
This research was supported by the AGH University  and  by the  Polish National Science Centre Grant: 2019/35/O/ST2/00357.

\bibliography{../../hydr.bib}

\end{document}